\documentclass[twocolumn]{revtex4-1}

\usepackage{amsmath}  
\usepackage{amsfonts,amssymb,bm} 
\usepackage{graphicx,color} 
\usepackage{xparse}
\usepackage{soul,physics}
\usepackage{hhline}
\usepackage{dsfont}
\usepackage[unicode=true, pdfusetitle, bookmarks=true, bookmarksnumbered=false, bookmarksopen=false, breaklinks=true, pdfborder={0 0 0}, backref=false, colorlinks=true, linkcolor=blue, citecolor=blue, urlcolor=blue]{hyperref}
\DeclareMathAlphabet{\mathbbold}{U}{bbold}{m}{n}
\DeclareRobustCommand{\Chi}{{\mathpalette\irchi\relax}}
\newcommand{\irchi}[2]{\raisebox{\depth}{$#1\chi$}}

\setstcolor{red}

\newcommand{\LLn}{$\text{L\'{e}vy-Leblond}$}
\newcommand{\LL}{$\text{L\'{e}vy-Leblond}$~}
\newcommand{\LLs}{$\text{L\'{e}vy-Leblond}$'s~}
\newcommand{\Schn}{$\text{Schr\"{o}dinger}$}
\newcommand{\Sch}{$\text{Schr\"{o}dinger}$~}

\newcommand{\SP}{$\text{Schr\"{o}dinger-Pauli}$~}
\newcommand{\ph}{{\phantom{-}}}
\newcommand{\vnab}{\grad}
\newcommand{\psid}{\psi^{\dagger}}

\begin{document}


\title{The Pauli and $\text{L\'{e}vy-Leblond}$ Equations, and the Spin Current Density}

\author{James M. Wilkes}
\email{mike.wilkes06@gmail.com} 
\altaffiliation[permanent address: ]{2127 NW Tilia Trail, Stuart, FL 34994, USA} %
\noaffiliation

\date{\today}

\begin{abstract}
We review the literature on the Pauli equation and its current density, discussing the progression from the original phenomenological version of Pauli to its derivation by \LL from a linearization of the \Sch equation. It was established conclusively by \LLs work that the spin of a spin-1/2 particle such as an electron is non-relativistic in nature, contrary to what was often stated following Dirac's derivation of a relativistic wave equation, and his subsequent demonstration that Pauli's spin interaction term appeared in the non-relativistic limit. In this limit, the Gordon decomposition of the associated probability current density was found to contain a spin-dependent term. Such a term does not follow, however, from the usual derivation of the current density from the Pauli equation, although various physically motivated but otherwise ad hoc explanations were put forward to account for it. We comment on the only exception to these of which we are aware implying the spin term in the current was in fact non-relativistic in nature. However, the earlier work of \LL had already shown, with no additional assumptions, that this term was a prominent feature of the current density derived from his equation. Hence, just as with the spin itself, the spin current was non-relativistic, claims to the contrary notwithstanding. We present a somewhat simplified derivation of the \LL equation and its current density, commenting on possibilities for experimental work that might indicate measurable consequences of the spin term in the current density.

\end{abstract}

\maketitle 

\section{Introduction}

Pauli's equation for the electron is a topic of general interest in quantum mechanics textbooks \cite{Levin,Greiner4,Ikenberry} and journal articles \cite{Galindo,Parker}. Although it explains many of the experimental results associated with non-relativistic electrons, its probability current density turns out to include an unexpected additional spin-dependent term (the spin current), which has received considerable attention  \cite{Parker,Nowakowski,Mita,Shikakhwa,Hodge}. We begin with a selective review of some of {the} relevant earlier work.

As background, in 1927 Pauli \cite{Pauli27} introduced his famous spin matrices in modifying the non-relativistic \Sch equation to account for Goudsmit and Uhlenbeck's  \cite{Goudsmit, Uhlenbeck} hypothesis that spectral data explainable by half-integral quantum numbers implied a {\emph {half-integer spin angular momentum}} for the electron. His ansatz was to add a phenomenological term to the usual Hamiltonian for an electron moving in an electromagnetic field, viz., the interaction energy of a magnetic field with an electron magnetic moment proportional to its intrinsic spin angular momentum. Introduction of spin matrices to describe this spin angular momentum necessitated replacing the complex scalar wave function of the \Sch equation by a two-component spinor wave function. The standard form of Pauli's equation as given in, for example, Ref.\ \onlinecite{Greiner4} (converted to SI units, used throughout the present work), is
\begin{align}
\left[\frac{1}{2m}\left(\hat{\bm{p}} -  q\bm{A}\right)^2 \mathds{1}+ q\phi \mathds{1}- \bm{\mu}_s\cdot \bm{B}\right]&\!\psi(\bm{r},t) = \nonumber\\ &\hat{E}\mathds{1}\,\psi(\bm{r},t)\,, \label{Paulieq}
\end{align}
where {$\psi= \begin{pmatrix}\psi_1\\\psi_2\end{pmatrix}$} is a 2-component spinor, $\mathds{1}$ is the $2\times 2$ identity matrix (often omitted, but to be understood), $m$ and $q$ \((= -e)\) are the mass and charge of the electron, and $\hat{\bm{p}} = -i\hbar \bm{\nabla}$ and $\hat{E} = i\hbar \frac{\partial}{\partial t}$ are the usual linear momentum and energy operators. The appearance of (real-valued) electromagnetic scalar and vector potentials, \ $\phi(\bm{r},t)$ and $\bm{A}(\bm{r},t)$, is a consequence of using the gauge invariant {\emph {minimal coupling}} assumption (see, for example, Ref. \onlinecite{Greiner4}, p. 205) to describe the interaction with external magnetic and electric fields defined by $\bm{B} = \bm{\nabla}\times \bm{A}$ and $\bm{E}=-\bm{\nabla}\phi - \frac{\partial}{\partial t}\bm{A}$. Finally,
\begin{align}
\bm{\mu}_s = \left(\frac{q\hbar}{2m}\right)\bm{\sigma} \label{emagmom}
\end{align}
is the electron magnetic moment operator, where $\bm{\sigma}$ is the Pauli spin vector with Pauli spin matrices $\sigma_i$, $i = 1, 2, 3$, as components.

The following year Dirac \cite{Dirac28} presented his relativistic wave equation for a free electron, and by making use of the minimal coupling replacement to include electromagnetic interactions showed that it contained two terms that ``can be regarded as the additional potential energy of the electron due to its new degree of freedom.'' One of these was, of course, the term involving the electron magnetic moment interacting with a magnetic field, introduced ``by hand'' in Pauli's equation. This prediction, requiring no ad hoc assumptions, was a triumph of the Dirac theory.

After Dirac's discovery it became common lore in many physics textbooks (see, for example, the extensive list of quotations compiled by \LL \cite{Levyleblond2}) to regard electron spin as a relativistic phenomenon, although there were prominent exceptions to this stance (Ref.\ \onlinecite{Levyleblond2}, p. 141).  Among these exceptions, e.g., Refs. \onlinecite{Galindo,Feynman,Sakurai,Adler}, the observation was made that the spin 1/2 property could be incorporated into the spin-0 non-relativistic \Sch equation by simply replacing the kinetic momentum operator $\hat{\bm{\pi}} = \hat{\bm{p}} - q \bm{A}$ appearing there with the product $\bm{\sigma}\cdot \hat{\bm{\pi}}$. To see how this is possible, we follow Sakurai \cite{Sakurai} by beginning with the spin-0 non-relativistic {\emph {free-electron}} \Sch equation of wave mechanics:
\begin{align}
\frac{\hat{\bm{p}}^2}{2m}\Psi(\bm{r},t) = \hat{E}\,\Psi(\bm{r},t)\,, \label{freeS}
\end{align}
where $\Psi$ is \Schn's complex scalar wave function. Recall the important spin-vector identity
\begin{align}
(\bm{\sigma}\cdot \hat{\bm{a}})(\bm{\sigma}\cdot \hat{\bm{b}}) = (\hat{\bm{a}}\cdot \hat{\bm{b}}) \mathds{1}+ i\,\bm{\sigma}\cdot (\hat{\bm{a}}\times \hat{\bm{b}})\,, \label{spinid}
\end{align}
where $\hat{\bm{a}}$ and $\hat{\bm{b}}$ are any two vector operators that commute with $\bm{\sigma}$. For $\hat{\bm{a}}=\hat{\bm{b}}=\hat{\bm{p}}$ one finds from Eq. \eqref{spinid} the operator identity $\left(\bm{\sigma}\cdot \hat{\bm{p}}\right)^2 = \hat{\bm{p}}^2 \mathds{1}$.\ Sakurai refers to the theory that results from making this substitution in \Schn's equation \eqref{freeS}, and replacing the wave function $\Psi$ by a 2-component spinor $\psi$, as the {\emph {\SP theory}} (Ref.\ \onlinecite{Sakurai}, footnote, p.\ 79), i.e., 
\begin{align}
\frac{\left(\bm{\sigma}\cdot \hat{\bm{p}}\right)^2}{2m}\psi(\bm{r},t) = \hat{E}\,\psi(\bm{r},t), \label{SP}
\end{align}
{which we refer to as the {\emph {\SP equation}}.} It has the plane wave solutions \(e^{i(\bm{k}\cdot\bm{r}-\omega t)}\,\Chi\), where \(\Chi\) is a constant but otherwise arbitrary {\emph {spinor}}, giving rise to the dispersion relation $\omega = \hbar\bm{k}^2/2m$ characteristic of non-relativistic particles.\ The effects of spin manifest by allowing the electron to interact with an electromagnetic field.\ This is accomplished by introduction of the gauge invariant minimal coupling replacements of $\hat{\bm{p}}$ with $\hat{\bm{\pi}} = \hat{\bm{p}} - q \bm{A}$, and $\hat{E}$ with $\hat{\varepsilon} = \hat{E} - q\phi$, in which case the {\SP equation} \eqref{SP} takes the form
\begin{align}
\frac{(\bm{\sigma}\cdot \hat{\bm{\pi}})^2}{2m}\,\psi(\bm{r},t) = \hat{\varepsilon}\kern0.04em\mathds{1}\,\psi(\bm{r},t). \label{MCSP}
\end{align}
Applying the identity \eqref{spinid} with $\hat{\bm{a}} = \hat{\bm{b}} = \hat{\bm{\pi}} =\,\,\hat{\bm{p}} - q \bm{A}$, we have in the numerator of the last line:
\begin{align}
(\bm{\sigma}\cdot \hat{\bm{\pi}})^2\psi &= \hat{\bm{\pi}}^2 \mathds{1}\psi + i\,\bm{\sigma}\cdot \left[\left(\hat{\bm{p}} - q \bm{A}\right)\times \left(\hat{\bm{p}} - q \bm{A}\right)\right]\psi\,,\notag\\
&= \hat{\bm{\pi}}^2\mathds{1}\psi - q\hbar\,\bm{\sigma}\cdot \left[\bm{\nabla}\times \left(\bm{A} \psi\right) + \bm{A}\times \left(\bm{\nabla}\psi\right) \right]\,,\notag\\
&= \hat{\bm{\pi}}^2\mathds{1}\psi - q\hbar\,(\bm{\sigma}\cdot \bm{B})\psi\,,\label{impid}
\end{align}
where the cross products $\hat{\bm{p}}\times \hat{\bm{p}}$ and $ \bm{A}\times\bm{A}$ both give vanishing contributions, we have replaced $\hat{\bm{p}} = -i\hbar\vnab$, and used the well-known vector identity $\vnab\times({\bm{A}\psi})=\vnab\psi\times\bm{A}+(\vnab\times\bm{A})\psi$, where $\vnab\times \bm{A} = \bm{B}$. Substituting Eq. \eqref{impid} together with $\hat{\varepsilon} = \hat{E} - q\phi$ in Eq. \eqref{MCSP} shows that the {\emph{minimally coupled}} \SP equation \eqref{MCSP} reproduces precisely the Pauli equation \eqref{Paulieq}. The result follows without regard for the Dirac equation, obtaining Pauli's spin interaction term directly from the minimally coupled \Sch equation and the replacement of $\hat{\bm{\pi}}$ by $\bm{\sigma}\cdot{\hat{\bm{\pi}}}$, as suggested by Feynman \cite{Feynman}.

The conclusion, then, is that the Pauli spin interaction term is an intrinsically non-relativistic effect. Still, one may be left with the feeling that if not for the purely mathematical and fortuitous existence of the spin identity, the Pauli equation would seem a slightly strange concoction forced upon the \Sch equation. This situation is somewhat alleviated by the fact that the minimally coupled \SP equation, from which Pauli's equation follows, can be shown \cite {Ikenberry, Sakurai, Bjorken} to be the {\emph{non-relativistic limit}} of Dirac's equation. Such a derivation is convincing, yet fails to satisfy the desire for a purely non-relativistic account. The work of \LLn, to be discussed later, seeks to fill this gap.

\section{The Missing Term in the Pauli Current Density}

As with the \Sch equation, one would hope to determine from the Pauli equation a probability density $\rho$ and current $\bm{J}'$ satisfying the probability conservation (or continuity) equation
\begin{align}
\frac{\partial \rho}{\partial t} + \bm{\nabla}\cdot \bm{J}' = 0\,. \label{probcon}
\end{align}
Such derivations are offered, for example, in Refs.\ \onlinecite{Levin}, \onlinecite{Greiner4}, and \onlinecite{Galindo}--\onlinecite{Mita}. A canonical derivation of the current density from Pauli's equation is that of Ref.\  \onlinecite{Greiner4}, {p. 340}. Defining the probability density by $\rho = \psid\psi$, where $\psid = \begin{pmatrix}\psi_1^* & \psi_2^*\end{pmatrix}$ denotes the Hermitian conjugate or adjoint of $\psi$, the current density is found there to be given (in our notation and SI units) by
\begin{align}
\bm{J}' = -\frac{i\hbar}{2m}\left[\psid\left(\vnab\psi\right) - \left(\vnab\psi\right)^\dagger\psi \right] - \frac{q}{m} \bm{A} \psid\psi\,. \label{PauliJG}
\end{align}

Since the Pauli equation is the non-relativistic limit of Dirac's equation, one might expect that Eq. \eqref{PauliJG} would be the corresponding non-relativistic limit of the current density of Dirac's equation, but such is not the case. Relativistic quantum mechanics shows that in this limit there appears an additional term, referred to as the spin current density. The details are given, for example, in Sakurai's discussion, Ref. \onlinecite{Sakurai}, p. 107, of the Gordon decomposition of the Dirac current and its non-relativistic limit. It is found (see also Ref.\ \onlinecite{Nowakowski}) that this limit yields instead the following expression:
\begin{align}
\bm{J} = \bm{J}' + \bm{J}_{{\textrm{spin}}}\,, \label{DiracNRJ}
\end{align}
where
\begin{align}
\bm{J}_{{\textrm{spin}}} \equiv \dfrac{\hbar}{2m}\bm{\nabla}\times\left(\psid\bm{\sigma}\psi\right) \label{JNRspin}
\end{align}
is the additional {\emph {spin current density}}. This term must be a feature of the current density for {\emph {any}} spin-1/2 particle described by the Dirac equation, even when the particle carries no charge, e.g., a neutron described by the Dirac equation with the addition of a non-minimal coupling term (as in Ref. \onlinecite{Bjorken}, p. 241). It is important to note, with Nowakowski \cite{Nowakowski}, that this spin current is also manifest in the reduction of the Dirac equation for a {\emph {free particle}}, i.e, in the {\emph {absence}} of any interactions.

Various writers have suggested methods for supplying the spin current missing from Eq.\ \eqref{PauliJG}. For example, in the venerable and oft-cited textbook of Landau and Lifshitz \cite {LandL},  the problem of a charged particle moving in a magnetic field is considered. A variational principle is applied, varying the expectation value of the Pauli Hamiltonian of Eq. \eqref{Paulieq} (ignoring the scalar potential) under a variation in the vector potential $\bm{A}$. From this they derive the {\emph {charge}} current density $q\bm{J}$, where $\bm{J}$ is given precisely by Eq. \eqref{DiracNRJ}, including the spin current. Without the minimal coupling to a magnetic field via $\bm{A}$ and the charge $q$, however, such a derivation apparently would fail to provide the desired spin current term, which we know from Eq. \eqref{DiracNRJ} should appear even in the absence of such an interaction.

Of other {works} known to the present author, each relies in some way upon the following line of reasoning. Since the continuity equation {\eqref{probcon}} involves only the divergence of the current density, it is always possible to add a divergenceless term to the current density without changing the equation. The spin current proportional to a curl is, of course, just such a term. But since the Pauli equation does not provide this, one must rely on reasonable {\emph {physical}} arguments to supply it, which should hold {\emph {even in the absence of {interaction potentials}}}.

As an example, Mita \cite {Mita} takes the novel approach of beginning with the expectation value of the spin angular momentum $\bm{S} = \frac{\hbar}{2}\,\bm{\sigma}$, and showing that it can be written as the volume integral of an angular momentum, viz.,
\begin{align}
\expval{\bm{S}} = m\int_{V_0}\bm{r}\times\bm{j}_s\,d^3{\text{r}}\,, ~ ~ \bm{j}_s = \frac{\hbar}{4m}\curl(\psi^{\dagger}\bm{\sigma}\psi)\,. \label{jmita}
\end{align}
Mita considers this $\bm{j}_s$ to be  ``the virtual probability current density that gives rise to the spin angular momentum of the electron.'' He also notes, however, that ``the proper gyromagnetic ratio of the particle cannot be obtained in the context of our analysis based on nonrelativistic quantum mechanics.'' In fact, his current density gives a gyromagnetic ratio that is half the correct value.

Greiner \cite{Greiner4} and Parker \cite{Parker}, as well as Hodge, et al \cite{Hodge} (following Greiner), call upon a result from electromagnetic theory, viz., that a magnetization field $\bm{M}$ induces an additional (electric) current density $q\bm{J}_M = \curl \bm{M}$ (see, for example, Refs.\ \onlinecite{GriffithsEM, Russakoff}). The expectation value of the electron magnetic moment is
\begin{align}
\expval{\bm{\mu}_s} = \int\psi^{\dagger}\bm{\mu}_s\psi~d^3{\text{r}} = \frac{q\hbar}{2m}\int\psi^{\dagger}\bm{\sigma}\psi~d^3{\text{r}}\,, \label{meanmu}
\end{align}
the integrand of Eq. \eqref{meanmu} defining the magnetic moment density, which these authors {\emph {identify}} with the magnetization $\bm{M}$ of electromagnetic theory. One thus finds for the associated probability current density:
\begin{align}
\bm{J}_M = \frac{1}{q}\curl\bm{M}  = \frac{\hbar}{2m}\curl\left(\psi^{\dagger}\bm{\sigma}\psi\right), \label{jhodge}
\end{align}
the correct expression for the spin current. As admitted in Ref.\  \onlinecite{Hodge}, referring to the current density, ``This argument is not rigorous, but it does hint that another term should be added $\dots$ .''

As a final example, we consider the work of Shikakhwa, {et al.} \cite{Shikakhwa}. They discuss Nowakowski's \cite{Nowakowski} spin current term, denoting it by $\bm{J}_M$ and referring to it as a magnetization current. These authors are clearly aware that the spin current cannot be derived from the Pauli equation \eqref{Paulieq} in the usual way (as, for example, in Ref.\  \onlinecite{Greiner4}, p. 340), so they instead begin with the minimally coupled \SP equation \eqref{MCSP}, ignoring the scalar potential term:
\begin{align}
\frac{\left(\bm{\sigma}\cdot \hat{\bm{\pi}}\right)^2}{2m}\psi = i\hbar\frac{\partial \psi}{\partial t}\,. \label{Shik}
\end{align}
The key to the success of their approach was to use Eq. \eqref{Shik} in deriving a probability conservation equation {\emph {before}} expanding $\left(\bm{\sigma}\cdot \hat{\bm{\pi}}\right)^2$ {(cf. Eq. \eqref{impid})}, since otherwise the Pauli equation \eqref{Paulieq} follows.\ Note that although their goal is to obtain Nowakowski's expression for the non-relativistic limit of the current density for a {\emph {free}} spin-1/2 particle, given in their Eq. (2), they actually work with Eq.\ \eqref{Shik}, above, for such a particle {\emph {interacting}} with a magnetic field. However, they emphasize more than once their use of $\hat{\bm{\pi}}$ as merely a convenient bookkeeping device in arriving at the desired result; the vector potential is ignored in the final step of their derivation. Their motivation for using $\hat{\bm{\pi}}$ rather than $\hat{\bm{p}}$, which naturally appears in the free particle Hamiltonian, is the concern that using $\hat{\bm{p}}$ in their derivation would be problematic due to the commutation of its components.\ We show in the next section that their concern can be addressed by a different derivation that avoids this problem.

In concluding this (in no way comprehensive) review, we are compelled to attempt clarification of certain assertions made in the Nowakowski \cite{Nowakowski} and Shikakhwa, {et al.} \cite{Shikakhwa} papers. In Ref.\ \onlinecite{Shikakhwa} one reads ``Nowakowski correctly states that the additional term $\bm{J}_M$ cannot be derived from the non-relativistic Pauli equation, because the covariance argument can be applied only for the fully relativistic Dirac equation.'' The first clause of this sentence appears to be based on Nowakowski's \cite
{Nowakowski} Eq.\ (4) for the current density derived from Pauli's equation, in which the spin current is notably absent. But this is already well known, and no surprise (e.g., as cited earlier in Ref.\  \onlinecite{Greiner4}, {p. 340}). The second clause refers to the  Lorentz covariance of the 4-vector probability current density of the Dirac equation, {whose form} can be shown \cite {Holland1, Holland2} to be {\emph {unique}}.

What the authors of Ref.\ \onlinecite{Shikakhwa} have shown, however, is that if one starts from the non-relativistic minimally coupled \SP equation of Eq.\ \eqref{Shik}, computing the current density before expanding the term $(\bm{\sigma}\cdot\bm{\pi})^2$, precisely the same form of the free particle current density found in the non-relativistic limit of the Dirac current density is obtained, including the correct spin current (after setting $\bm{A}$ {to zero}). Contrary to what Nowakowski \cite{Nowakowski} suggests, their work shows (although they seem reluctant to make this claim) that there is {\emph {no ambiguity}} in the definition of the current density derived from the non-relativistic \SP equation.\ True enough, the uniqueness of the non-relativistic current density was not discussed there, but in fact its uniqueness was found in earlier work of Holland \cite{Holland1, Holland2} and Holland and Philippidis \cite{Holland3} to be {\emph {inherited}} as the limit of the unique current density of the relativistic case.  The situation is reminiscent of the claim by early proponents of the Dirac equation that relativity was necessary to explain the electron's spin. The claim that relativity is required to explain the {\emph {spin current}} is similarly unfounded. We believe that Shikakhwa, {et al.} \cite{Shikakhwa} have shown it to be derivable from the non-relativistic \SP equation \eqref{Shik}, although just as with the spin, relativity proves to be invaluable in confirming its existence and uniqueness.

\section{Derivation of the Spin Current Density from the Schr\"{o}dinger-Pauli Equation for a Free Spin-1/2 Particle}\label{free}

The authors of Ref. \onlinecite{Shikakhwa} derived the non-relativistic current density for a free spin-1/2 particle beginning with the minimally coupled \SP equation, and later set $\bm{A} = \bm{0}$ in the result. We show here that the current density can be derived directly from the free-particle \SP equation \eqref{SP}. From this equation we have
\begin{align*}
\frac{\partial \psi}{\partial t} = -\frac{i}{2m\hbar}\left(\bm{\sigma}\cdot \hat{\bm{p}}\right)^2\psi\,,
\end{align*}
and its Hermitian conjugate equation:
\begin{align*}
\frac{\partial \psi^{\dagger}}{\partial t} = \frac{i}{2m\hbar}\left[\left(\bm{\sigma}\cdot \hat{\bm{p}}\right)^2\psi\right]^{\dagger}.
\end{align*}
Taking $\rho = \psi^{\dagger}\psi$ as the probability density, we find for its time derivative
\begin{align}
\frac{\partial \rho}{\partial t} &= \psi^{\dagger}\frac{\partial \psi}{\partial t} + \frac{\partial \psi^{\dagger}}{\partial t}\psi\,,\notag\\
&= -\frac{i}{2m\hbar} \left\{\psi^{\dagger}\!\left[\left(\bm{\sigma}\cdot \hat{\bm{p}}\right)^2\psi\right] - \left[\left(\bm{\sigma}\cdot \hat{\bm{p}}\right)^2\psi\right]^{\dagger}\psi\right\}. \label{firstrhot}
\end{align}
At this point, our derivation departs from the one given in Ref. \onlinecite{Shikakhwa}, where our momentum operator $\hat{\bm{p}}$ in Eq. \eqref{firstrhot} was replaced by the kinetic momentum $\hat{\bm{\pi}}$ in their corresponding Eq. (8b).\ We instead retain $\hat{\bm{p}}$ and introduce an auxiliary spinor $\Chi$ defined by
\begin{align}
\Chi = -\frac{1}{2mc}\left(\bm{\sigma}\cdot \hat{\bm{p}}\right)\psi\,, \label{chi}
\end{align}
the speed of light $c$ included only to ensure dimensional consistency (in no way to be construed as introducing a relativistic concept into the theory). Eq. \eqref{firstrhot} can then be written as
\begin{align*}
\frac{1}{c}\frac{\partial \rho}{\partial t} &= \frac{i}{\hbar}\,\psi^{\dagger}\big[\!\left(\bm{\sigma}\cdot \hat{\bm{p}}\right)\Chi\big] - \frac{i}{\hbar}\,\big[\!\left(\bm{\sigma}\cdot \hat{\bm{p}}\right)\Chi\big]^{\dagger}\psi\,, \\
&= \psi^{\dagger}\left(\bm{\sigma}\cdot{\bm{\nabla}}\Chi\right) + \left({\bm{\nabla}}\Chi\cdot{\bm{\sigma}}\right)^{\dagger}\psi\,.
\end{align*}
The divergence identity
\begin{align*}
\div (\psi^{\dagger}\bm{\sigma}\Chi + \Chi^{\dagger}\bm{\sigma}\psi) &= \, \psi^{\dagger}\left(\bm{\sigma}\cdot{\bm{\nabla}}\Chi\right) + \left({\bm{\nabla}}\Chi\cdot{\bm{\sigma}}\right)^{\dagger}\psi\\ &+  \, \left({\bm{\nabla}}\Chi\cdot{\bm{\sigma}}\right)^{\dagger}\Chi + \Chi^{\dagger}(\bm{\sigma}\cdot \vnab \psi)\,,
\end{align*}
allows us to replace the right-hand side of the last equation, yielding
\begin{align*}
\frac{1}{c}\frac{\partial \rho}{\partial t} = \div (\psi^{\dagger}\bm{\sigma}\Chi + \Chi^{\dagger}\bm{\sigma}\psi)- \left({\bm{\nabla}}\Chi\cdot{\bm{\sigma}}\right)^{\dagger}\Chi  - \Chi^{\dagger}(\bm{\sigma}\cdot \vnab \psi)\,.
\end{align*}
From equation \eqref{chi} and its adjoint, however, we have
\begin{align*}
\Chi = \dfrac{i\hbar}{2mc}\bm{\sigma}\cdot {\bm{\nabla}}\psi\,, ~ ~ ~
\Chi^{\dagger} = -\dfrac{i\hbar}{2mc}\left(\bm{\sigma}\cdot {\bm{\nabla}}\psi\right)^{\dagger}\,,
\end{align*}
which, when substituted in the previous equation eliminates the last two terms, bringing it to simply
\begin{align}
\frac{1}{c}\frac{\partial \rho}{\partial t} = \vnab \cdot (\psi^{\dagger}\bm{\sigma}\Chi + \Chi^{\dagger}\bm{\sigma}\psi)\,. \label{ptrho1}
\end{align}
Substituting for $\Chi$ and $\Chi^{\dagger}$ in \eqref{ptrho1} (thereby {\emph {eliminating}} the dimensional factor of $c$) yields
\begin{align}
\frac{\partial \rho}{\partial t} &= \dfrac{i\hbar}{2m}\left\{\vnab\cdot\left[\psi^{\dagger}\bm{\sigma}(\bm{\sigma}\cdot \vnab \psi) - (\vnab \psi\cdot \bm{\sigma})^\dagger\bm{\sigma}\psi\right]\right\}\,,\nonumber\\ &= -\div \bm{J}\,, \label{J1}
\end{align}
allowing identification of the components of the probability current density $\bm{J}$:
\begin{align*}
J_i = \dfrac{i\hbar}{2m}\big[(\nabla_j \psi)^{\dagger} \sigma_j\sigma_i\psi - \psi^{\dagger}\sigma_i\sigma_j(\nabla_j \psi) \big]\,,
\end{align*}
summation over repeated indices being understood. Using the Pauli matrix identity \(\sigma_i\sigma_j = \delta_{ij} + i\epsilon_{ijk}\sigma_k\), the current density components can be written as
\begin{align*}
J_i &= \dfrac{i\hbar}{2m}\left((\nabla_j \psi)^{\dagger} \left(\delta_{ij} - i\epsilon_{ijk}\sigma_k\right)\psi - \psi^{\dagger}\left(\delta_{ij} + i\epsilon_{ijk}\sigma_k\right)\nabla_j \psi\right)\,,\\
&= \dfrac{i\hbar}{2m}\left[(\nabla_i \psi)^{\dagger})\psi - \psi^{\dagger}(\nabla_i \psi) - i\epsilon_{ijk}\nabla_j(\psi^{\dagger}\sigma_k\psi)\right]\,,
\end{align*}
or in vector form:
\begin{align}
\bm{J} = -\dfrac{i\hbar}{2m}&\left[\psi^{\dagger}(\vnab \psi) - (\vnab \psi)^{\dagger}\psi\right]\nonumber\\ &+ \dfrac{\hbar}{2m}\left[\vnab \times (\psi^{\dagger}\bm{\sigma}\psi) \right], \label{Jtotal}
\end{align}
the correct current density for an external interaction-free spin-1/2 particle.

\section{The L\'{e}vy-Leblond Equation}

Nearly 40 years after Pauli and Dirac's seminal work, in a paper \cite{Levyleblond1} ``devoted to a detailed study of non-relativistic particles and their properties, as described by Galilei invariant wave equations,'' J.-M. L\'{e}vy-Leblond, inspired by Dirac's heuristic derivation \cite{Dirac28} of his relativistic wave equation for the electron, obtained from the free-particle \Sch equation a non-relativistic analog now referred to as the L\'{e}vy-Leblond equation. From it the Pauli equation {\emph {follows}} as an immediate consequence after calling upon the minimal coupling replacement. In addition, the probability current computed from the \LL equation includes the correct spin current contribution.

A derivation of the \LL equation eventually involves the introduction of a 4-component spinor (bispinor), and $4\times 4$ matrices satisfying the Dirac algebra, which may be considered at too high a level to be introduced in an early course on quantum mechanics. As we show, however, these matrices are easily constructed from the Pauli matrices, so if the Pauli theory has been introduced, there is really no great hurdle to be overcome in using them. Even so, if the decision is made to avoid their use, we should mention that having made plausible the Pauli equation by, e.g., the Sakurai argument leading to Eq. \eqref{MCSP}, a coupled set of equations \emph{linear} in the momentum operator, i.e., first order in the spatial derivatives, can be obtained by simply introducing, similar to what was done in the last section, an {\emph {auxiliary}} spinor $\Chi$ defined by
\begin{align}
\Chi = -\frac{1}{2mc}\left[\bm{\sigma}\cdot \left(\hat{\bm{p}} - q \bm{A}\right)\right]\,\psi\,, \label{chidef}
\end{align}
It then follows from Eq. \eqref{MCSP} that
\begin{align}
-c\left[\bm{\sigma}\cdot \left(\hat{\bm{p}} - q \bm{A}\right)\right]\,\Chi + q\phi\,\psi = \hat{E}\,\psi. \label{LLdef}
\end{align}
This {\emph {pair}} of equations is precisely the system derived by \LL in Ref.\ \onlinecite{Levyleblond1}, that is, the \LL equations {\emph {follow}} from the Pauli equation. \LLs distinguished contribution, on the other hand, was to provide a {\emph {derivation}} of these equations from the non-relativistic \Sch equation, from which Pauli's equation then {\emph {follows}}. A derivation based on his insights is offered in what follows.

\subsection{Linearizing the \Sch Equation}

We begin with the \Sch equation \eqref{freeS} for a free particle of mass $m$, multiplying both sides by $2m$ to write it in the more convenient form 
\begin{align}
\left(\hat{p}_i\hat{p}_j\delta_{ij} - 2m\hat{E}\right)\Psi(\bm{r},t) =  0. \label{Sch}
\end{align}
Following \LL \cite{Levyleblond1}, we seek a wave equation linear in both the energy and momenta operators:
\begin{align}
\hat{\theta}\,\Psi(\bm{r},t) \equiv \left(\frac{\hat{A}}{c}\hat{E} + \hat{B}_i\hat{p}_j\delta_{ij} + mc\,\hat{C}\right)\!\Psi(\bm{r},t) =  0\,, \label{linS}
\end{align}
where $\hat{A}$, $\hat{C}$, and $\hat{B}_i ~ (i = 1, 2, 3)$ are dimensionless linear operators to be determined. They are assumed to depend on neither the spatial coordinates nor time, hence commute with $\hat{E}$ and each $\hat{p}_i$, but not necessarily with each other. The factors $1/c$ of $\hat{A}$, and $mc$ of $\hat{B}$, where $c$ is an arbitrary constant speed, are introduced to insure the correct units of linear momentum in each term. It will later be identified with the speed of light for compatibility with the non-relativistic reduction of the Dirac equation. From Eq. \eqref{linS} we obtain, by operating on both sides with $\hat{\theta}$,
\begin{align}
\hat{\theta}^2\,\Psi(\bm{r},t) \equiv \left(\frac{\hat{A}}{c}\hat{E} + \hat{B}_i\hat{p}_j\delta_{ij} + mc\,\hat{C}\right)^{\!\!2}\Psi(\bm{r},t) =  0\,. \label{thetasq}
\end{align}
In doing so we have departed from \LLs \cite{Levyleblond1} and other derivations (e.g., Ref.\ \onlinecite{Greiner4}, Ch.\ 13, and Refs.\ \cite{Dutoit, Hladik}) by omitting {\emph{primed counterparts}} of the operators $\hat{A}$, $\hat{B}_i$, and $\hat{C}$. Conditions are now set on these operators by requiring that Eqs. \eqref{Sch} and \eqref{thetasq} agree. Squaring the linear operator, we obtain
\begin{align*}
\frac{\hat{A}^2}{c^2}\,\hat{E}^2 &+ \frac{1}{c}(\hat{A}\hat{B}_i + \hat{B}_i\hat{A})\hat{E}\hat{p}_i + m(\hat{A}\hat{C} + \hat{C}\hat{A})\hat{E} \\& + \frac{1}{2}(\hat{B}_i\hat{B}_j + \hat{B}_j\hat{B}_i)\hat{p}_i\hat{p}_j + mc (\hat{B}_i\hat{C} + \hat{C}\hat{B}_i)\hat{p}_i\qquad\\  &+ m^2c^2\,\hat{C}^2 \,=\, \hat{p}_i\hat{p}_j\delta_{ij} - 2m\hat{E},
\end{align*}
where the fourth term on the left-hand side was obtained by interchanging summation indices to rewrite it as a symmetrized sum. This polynomial equation in $\hat{E}$ and the $\hat{p}_i$ is satisfied for all $\hat{A}$, $\hat{C}$, and $\hat{B}_i$ if and only if each of its coefficients is zero, yielding
\begin{gather}
\hat{A}^2 = \hat{0}\,, ~ ~ \hat{C}^2 = \hat{0}\,, ~ ~ \hat{A}\hat{C} + \hat{C}\hat{A} = -2\,\hat{I}\,, \notag\\
\hat{A}\hat{B}_i + \hat{B}_i\hat{A} = \hat{0}\,, ~ ~ \hat{C}\hat{B}_i + \hat{B}_i\hat{C} = \hat{0}\,, \label{ABCconds}\\
\hat{B}_i\hat{B}_j + \hat{B}_j\hat{B}_i = 2\delta_{ij}\,\hat{I}\,,~ ~ (i, j = 1, 2, 3)\,. \notag
\end{gather}
where $\hat{I}$ is the identity operator. 

Along with the operators $\hat{B}_i$, whose anticommutation relations are satisfied by the three Pauli {\emph {spin matrix representations}} $B_i = \sigma_i$ (the ``hat'' will be omitted in denoting matrix representations of operators), two additional operators $\hat{A}$ and $\hat{C}$ have been introduced. We see in Eqs. \eqref{ABCconds} that the matrix representations of both operators must anticommute with each of the ${B_i}$ matrix representations, but unlike the Pauli matrices, they neither anticommute with each other nor does each square to the identity matrix (both are in fact singular). Noticing that anticommutation with the ${B_i}$ is maintained when the representations ${A}$ and ${C}$ are each replaced by any linear combination of the two, we seek such linear combinations, ${B}_4$ and ${B}_5$, having the same anticommutation relations as the Pauli matrices in order to take advantage of their well documented properties. It is not difficult to demonstrate that they must have the general forms, up to a choice of sign on $i$,
\begin{align}
B_4 = \frac{1}{2}\left(\frac{A}{a} + \frac{C}{b}\right)\,, ~~~
B_5 = -\frac{i}{2}\left(\frac{A}{a} - \frac{C}{b}\right)\,, \label{B4B5def}
\end{align}
provided the nonzero dimensionless constants $a$ and $b$ satisfy
\begin{equation}
2ab = -1  \label{abdef}\,.
\end{equation} 
That is, when $A$, $C$, and the $B_i$ satisfy Eqs. \eqref{ABCconds}, then Eqs. \eqref{B4B5def} and \eqref{abdef} insure that the five matrices $B_\mu$, $\mu = 1, \ldots, 5$, satisfy the anticommutation relations
\begin{align}
B_\mu B_\nu + B_\nu B_\mu = 2\delta_{\mu\nu}\,I\,, ~ ~ ~ ~ \mu, \nu = 1, \ldots, 5\,. \label{Diracalg5}
\end{align}

Our situation is reminiscent of Dirac's derivation of his relativistic wave equation, where he found that the three $2\times 2$ Pauli matrices were not sufficient to satisfy his anticommutation relations. It was necessary to generalize the three Pauli matrices to four $4\times 4$ matrices $\alpha_p$, $p = 1, \ldots, 4$, for which
\begin{align}
\alpha_p \alpha_q + \alpha_q \alpha_p = 2\delta_{pq}\,I, \qquad p, q = 1,
\ldots 4, \label{Dirac4}
\end{align} 
where $I$ is the $4\times 4$ identity matrix (a demonstration of the necessity for $4\times 4$ matrices is given in, for example, Ref.\ \onlinecite{Bjorken}, p. 8). We see now, in light of Eq. \eqref{Diracalg5}, that our proposed linearization of the \Sch equation can be completed by finding {\emph {five}} complex $4\times 4$ matrices satisfying Dirac's algebra. The defining equations \eqref{B4B5def} are then easily inverted to get expressions for the original matrices $A$ and $C$ of Eq. \eqref{linS}, viz.,
\begin{align}
A = a(B_4 + iB_5)\,, ~~~ 
C = b(B_4 - iB_5)\,. \label{ACmats}
\end{align}

Fortunately, if four initial matrices satisfying Eq.\ \eqref{Dirac4} are {\emph {known}}, then the matrix $B_5 = B_1B_2B_3B_4$ is the desired fifth matrix, i.e., a calculation using the properties of the $B_p$ matrices, $p = 1, \dots, 4$, shows that $B_5^2 = I$, and $B_5B_p = - B_p B_5$ for each $p$ taken separately.\ For example, the demonstration that $B_5^2 = I$ can be accomplished in three steps: (1) focusing on the first factor of $B_5$ in the square, use the anticommutation relations to move $B_1$ three places to the right, and replace $B_1^2 = I$, (2)~move $B_4$ two places to the right, and replace $B_4^2 = I$, and (3) commute $B_3$ with $B_2$, replacing $B_2^2 = B_3^2 = I$ in the result. Each commutation changes the sign on the product, so we have 
\begin{align*}
    B_5^2 &= (B_1B_2B_3B_4)(B_1B_2B_3B_4),\\
    &= -B_2B_3B_4B_1^2B_2B_3B_4,\\
    &= -B_2B_3B_4B_2B_3B_4,\\
    &= -B_2B_3B_2B_3B_4^2,\\
    &= -B_2B_3B_2B_3,\\
    &= {\phantom{-}}B_2^2B_3^2 \,=\, I\,.
\end{align*} 
By similar manipulations, one finds that $B_5B_p = - B_p B_5$ for each $p$ taken separately.

We emphasize that our matrices $B_\mu$, $\mu = 1, \,\ldots, 5$ are \emph{not} the same as those introduced by either \LL in Ref.\ \onlinecite{Levyleblond1}, Greiner in Ref.\ \onlinecite{Greiner4} (Ch.\ 13), Hladik \cite{Hladik}, or Du Toit \cite{Dutoit}. In all four references their matrices are intimately tied to primed counterparts that we have not introduced (and, incidentally, are different in each reference), and {\emph {do not}} satisfy the anticommutation relations \eqref{Diracalg5} of the Dirac algebra.

\subsection{Constructing Representations of the Dirac Algebra}

The first four of our five matrices could be chosen to be Dirac's original matrices, i.e., $B_p = \alpha_p$, $p = 1, \ldots, 4$. They can be constructed as Kronecker products of Pauli matrices \cite{Arfken3rd, Snygg, Goodmanson}, where the Kronecker product of two $2\times 2$ matrices $M$ and $N$ is a $4\times 4$ matrix defined by
\begin{align}
    M\otimes N &= \begin{pmatrix}m_{11} & m_{12}\\m_{21} & m_{22}\end{pmatrix}\otimes N
= \begin{pmatrix}m_{11}N & m_{12}N\\m_{21}N & m_{22}N\end{pmatrix} \nonumber\,,\\[2pt]
&= \begin{pmatrix}m_{11}n_{11} & m_{11}n_{12} & m_{12}n_{11} & m_{12}n_{12}\\ m_{11}n_{21} & m_{11}n_{22} & m_{12}n_{21} & m_{12}n_{22}\\
m_{21}n_{11} & m_{21}n_{12} & m_{22}n_{11} & m_{22}n_{12}\\
m_{21}n_{21} & m_{21}n_{22} & m_{22}n_{21} & m_{22}n_{22}\end{pmatrix}. \label{Kron}
\end{align}
Recalling the Pauli matrices in their standard representation,
\begin{align}
\sigma_1 = \begin{pmatrix}0 & 1\\1 & 0\end{pmatrix}, \quad\sigma_2 = \begin{pmatrix}0 & -i\\i & \ph 0\end{pmatrix}, \quad\sigma_3 = \begin{pmatrix}1 & \ph 0\\0 & -1\end{pmatrix},\label{Paulis}
\end{align}
Dirac's original matrices are the following Kronecker products ($i = 1, 2, 3$):
\begin{subequations}\label{alphas}
\begin{align}
\alpha_i &= \sigma_1\otimes \sigma_i = \begin{pmatrix}0 & 1\\1 & 0\end{pmatrix}\otimes \sigma_i = \begin{pmatrix}\mathbbold{0} & \sigma_i\\\sigma_i & \mathbbold{0}\end{pmatrix}, \label{alphai}\\[3pt]
\beta &= \alpha_4 = \sigma_3\otimes \mathds{1}= \begin{pmatrix}1 & \ph 0\\0 & -1\end{pmatrix}\otimes \mathds{1}= \begin{pmatrix}\mathds{1}& \ph\mathbbold{0}\\\mathbbold{0} & -\mathds{1}\end{pmatrix},\label{alpha4}
\end{align}
\end{subequations}
where $\mathbbold{0}$ is the $2\times 2$ zero matrix, and $\mathds{1}$ is the $2\times 2$ identity matrix introduced earlier. A fifth matrix of the Dirac algebra is then given by
\begin{align}
B_5 = \alpha_1\alpha_2\alpha_3\alpha_4 = \begin{pmatrix}\mathbbold{0} & -i\mathds{1}\\i\mathds{1}& \ph\mathbbold{0} \end{pmatrix},\label{B5Kron}
\end{align}
using the Pauli spin matrix identity $\sigma_1\sigma_2\sigma_3 = i \mathds{1}$.\ One easily verifies that any pair of the five matrices $B_i = \alpha_i$, $i = 1, 2, 3$, $\,B_4 = \alpha_4$, and $B_5$ satisfies Eqs. \eqref{Diracalg5}.

A more convenient set of matrices, however, is obtained from the following Kronecker products ($i = 1, 2, 3$):
\begin{subequations}\label{Bconv}
\begin{align}
B_i &=  \ph\sigma_3\otimes \sigma_i = \begin{pmatrix}1 & \ph 0\\0 & -1\end{pmatrix}\otimes \sigma_i = \begin{pmatrix} ~ \sigma_i & \mathbbold{0}\\\mathbbold{0} & -\sigma_i ~ \end{pmatrix},\label{Biconv}\\
B_4 &= \ph\sigma_1\otimes \mathds{1} = \begin{pmatrix}0 & 1\\1 & 0\end{pmatrix}\otimes \mathds{1}= \begin{pmatrix}\mathbbold{0} & \mathds{1}\\\mathds{1}& \mathbbold{0} \end{pmatrix},\label{B4conv}
\end{align}
in which case the fifth matrix is
\begin{align}
B_5 = B_1B_2B_3B_4 =  \begin{pmatrix}\ph\mathbbold{0} & i\mathds{1}\\-i\mathds{1}& \,\mathbbold{0} \end{pmatrix}.\label{B5conv}
\end{align}
\end{subequations}
These matrices also satisfy Eqs.\ \eqref{Diracalg5}, and with their choice the matrices $A$ and $C$ of Eqs. \eqref{ACmats} take the simple forms
\begin{align}
A = 2a\begin{pmatrix}\mathbbold{0} & \mathbbold{0}\\ \mathds{1}&\mathbbold{0} \end{pmatrix}, \qquad
C = 2b\begin{pmatrix}\mathbbold{0} & \mathds{1}\\\mathbbold{0} & \mathbbold{0} \end{pmatrix}. \label{ACconv}
\end{align}

\subsection{The Levy-Leblond and Pauli Equations}

In the representations of Eqs. \eqref{Bconv} and \eqref{ACconv} the linearized \Sch equation \eqref{linS} takes the matrix form
\begin{align}
    &\big(\frac{A}{c}\,\hat{E} + {B}_i\hat{p}_j\delta_{ij} + mc\,{C}\big)\Psi\nonumber  \\[3pt]
    & =\left[\dfrac{2a}{c}\begin{pmatrix}\mathbbold{0} & \mathbbold{0}\\\mathds{1}&\mathbbold{0} \end{pmatrix}\hat{E} + \begin{pmatrix} ~ \sigma_i & \ph\mathbbold{0}\\\mathbbold{0} & -\sigma_i ~ \end{pmatrix}\hat{p}_i + 2bmc\begin{pmatrix}\mathbbold{0} & \mathds{1}\\\mathbbold{0} & \mathbbold{0} \end{pmatrix}\right]\!\Psi \nonumber \,, \\[5pt]
    &=\dfrac{1}{c}\!\begin{pmatrix} ~ ~ c\,\bm{\sigma}\cdot \hat{\bm{p}} & ~ ~ ~ ~ 2bmc^2\mathds{1}~ ~ \\2a\mathds{1}\hat{E} & ~ -c\,\bm{\sigma}\cdot \hat{\bm{p}} ~ ~ \end{pmatrix}\!\Psi = \bm{0}\,,\label{LLeq}
\end{align}
%
where we have omitted the position and time arguments of $\Psi$ for convenience, and note that $\Psi$ must now be a bispinor, i.e., a $4$-element column matrix of complex functions (\(\bm{0}\) is the zero bispinor). Eq.\ \eqref{LLeq} is the {\emph {\LL equation}} for a free particle of mass $m$. Various forms of it can be found in the literature, depending on the choices made for $a$ and $b$ subject to condition \eqref{abdef}. We write the bispinor $\Psi$ as
\begin{align}
\Psi=\begin{pmatrix}\psi \\ \Chi \end{pmatrix}, \label{Phi}
\end{align}
where $\psi$ and $\Chi$ are two-component spinors (sometimes referred to as semispinors), to obtain the component equations
\begin{subequations}
\begin{align*}
\bm{\sigma}\cdot \hat{\bm{p}}\,\psi + 2bmc\mathds{1}\Chi &= \bm{0}\, \\
-2a\mathds{1}\hat{E}\,\psi + c\bm{\sigma}\cdot \hat{\bm{p}}\,\Chi &= \bm{0}\,.
\end{align*}
\end{subequations}
When we choose $a = -1/2$, hence $b = 1$, these reduce to \LLs Eqs.\ (25) of Ref.\ \onlinecite{Levyleblond1}, up to missing factors of $c$ required for correct SI units. Both Greiner's (Ref.\ \onlinecite{Greiner4}, Eqs.\ (13.28)), and Hladik's (Ref.\ \onlinecite{Hladik}), versions of the \LL equations correspond to the choice $a= i/2$, hence $b= i$, while Du Toit's equations (Ref.\ \onlinecite{Dutoit}) correspond to $a= 1/2$, hence $b = -1$. In all that follows, we adhere to \LLs original equations (providing the missing factors of $c$), hence
\begin{subequations}\label{LL1comps}
\begin{align}
\bm{\sigma}\cdot \hat{\bm{p}}\,\psi + 2mc\mathds{1}\Chi &= \bm{0}\,,\label{comp1}\\
\mathds{1}\hat{E}\,\psi + c\bm{\sigma}\cdot \hat{\bm{p}}\,\Chi &= \bm{0}\,,\label{comp2}
\end{align}
\end{subequations}
in agreement with the non-relativistic limit of the Dirac equations if $c$ is identified with the speed of light.

If the particle has a charge $q$ and is moving in an external electromagnetic field derived from a vector potential $\bm{A}$ and scalar potential $\phi$, then Eqs. \eqref{LL1comps} are modified to include the interaction using the minimal coupling prescription, as was done in Eq. \eqref{MCSP}:
\begin{subequations}\label{LLinter}
\begin{align}
\bm{\sigma}\cdot \left(\hat{\bm{p}} -  q\bm{A}\right)\psi + 2mc\mathds{1}\,\Chi &= \bm{0}\,,\label{inter1}\\
c\bm{\sigma}\cdot \left(\hat{\bm{p}} -  q\bm{A}\right)\Chi +(\hat{{E}} - q\phi)\mathds{1}\,\psi&= \bm{0}\,.\label{inter2}
\end{align}
\end{subequations}
Eliminating $\Chi$ between the two equations by multiplying Eq. \eqref{inter2} by $2m$, then using Eq. \eqref{inter1} to replace $2mc\Chi$ in the result, we obtain a wave equation for the spinor $\psi$:
\begin{align}
\big[\bm{\sigma}\cdot \left(\hat{\bm{p}} -  q\bm{A}\right)\big]^2\,\psi - 2m (\hat{{E}} - q\phi)\mathds{1}\,\psi = \bm{0}\,,\label{Pauli1}
\end{align}
the factors of $c$ obligingly falling out (showing once again that special relativity is not involved). Dividing both sides by $2m$, we see that this is precisely the minimally coupled \SP equation, Eq. \eqref{MCSP}, from which we have shown that Pauli's equation \eqref{Paulieq} follows directly. Thus, as emphasized by \LL in Ref.\ \onlinecite{Levyleblond1}, the linearized {\emph {non-relativistic}} (Galilean covariant) \Sch equation together with minimal coupling predicts the correct value for the intrinsic magnetic moment of a spin-1/2 particle: spin is {\emph {not}} an intrinsically relativistic phenomena.

\section{Derivation of the Spin Current Density from the L\'{e}vy-Leblond Equation}

It seems not well known, although \LL published it over 40 years ago, that his equation predicts unambiguously the form of the spin current showing that, like the intrinsic spin itself, it is an inherently non-relativistic phenomena.

To demonstrate this we replace the energy operator in the components, Eqs. \eqref{LLinter}, of the \LL equation, yielding
\begin{subequations}\label{LLops}
\begin{align}
\frac{\partial \psi}{\partial t} &=  \frac{ic}{\hbar}\bm{\sigma}\cdot \left(\hat{\bm{p}} -  q\bm{A}\right)\Chi - \frac{i}{\hbar}q\phi\mathds{1} \psi\,,\label{ops1}\\
\Chi &= -\frac{1}{2mc}\bm{\sigma}\cdot \left(\hat{\bm{p}} -  q\bm{A}\right)\psi\,,\label{ops2}
\end{align}
\end{subequations}
obvious generalizations of the free particle equations introduced as an {\emph {ansatz}} in section \ref{free}. In this form we see that $\Chi$ plays an auxiliary role, as it is given in terms of the component $\psi$ through the constraint \eqref{ops2}. On the other hand, Eq. \eqref{ops1} is a true dynamical equation describing the time evolution of the state \(\Psi\) via its independent component \(\psi\). Defining the probability density again by $\rho = \psi^{\dagger}\psi$, we have
\begin{align}
\frac{\partial \rho}{\partial t} &= \psi^{\dagger}\frac{\partial \psi}{\partial t} + \frac{\partial \psi^{\dagger}}{\partial t}\psi\,. \label{dender}
\end{align}
The adjoint of \eqref{ops1} is
\begin{align}
\frac{\partial \psi^\dagger}{\partial t} &=  -\frac{ic}{\hbar}\big[\bm{\sigma}\cdot \left(\hat{\bm{p}} -  q\bm{A}\right)\, \Chi\big]^\dagger + \frac{i}{\hbar}q\phi \,\psi^{\dagger}\mathds{1}\,,\label{ops1adj}
\end{align}
hence from Eqs.\ \eqref{ops1}, \eqref{ops1adj}, and \eqref{dender}:
\begin{align*}
    \frac{1}{c}\frac{\partial \rho}{\partial t} &= \frac{i}{\hbar}\Big\{\psi^{\dagger}\big[\bm{\sigma}\cdot\left(\hat{\bm{p}} -  q\bm{A}\right)\Chi\big] - \big[\bm{\sigma}\cdot\left(\hat{\bm{p}} -  q\bm{A}\right)\Chi\big]^{\dagger}\psi\Big\}, \\
    &= \psi^{\dagger}\left(\bm{\sigma}\cdot{\bm{\nabla}}\Chi\right) + (\bm{\sigma}\cdot\grad \Chi)^{\dagger}\psi- \frac{i q}{\hbar}\Big\{\psi^{\dagger}\big[\!\left(\bm{\sigma}\cdot\bm{A}\right)\Chi\big]\\
    &\kern4.8cm- \big[\!\left(\bm{\sigma}\cdot \bm{A}\right)\Chi\big]^{\dagger}\psi\Big\},
\end{align*}
%
the terms involving $q\phi$ cancelling out. The first two terms can be rewritten in terms of a divergence, to obtain
\begin{align}
\frac{1}{c}\frac{\partial \rho}{\partial t} &= \div (\psi^{\dagger}\bm{\sigma}\Chi + \Chi^{\dagger}\bm{\sigma}\psi) - (\bm{\sigma}\cdot\grad \psi)^{\dagger}\Chi - \Chi^{\dagger}(\bm{\sigma}\cdot \vnab \psi) \nonumber\\
 &\quad- \frac{iq}{\hbar}\Big\{\psi^{\dagger}\big[\!\left(\bm{\sigma}\cdot\bm{A}\right)\Chi\big] - \big[\!\left(\bm{\sigma}\cdot \bm{A}\right)\Chi\big]^{\dagger}\psi\Big\}.\label{ptrhoA1}
\end{align}
From Eq.\ \eqref{ops2} and its adjoint, however, we have
\begin{align*}
\Chi &= -\frac{1}{2mc}\big[\bm{\sigma}\cdot \left(\hat{\bm{p}} -  q\bm{A}\right)\, \psi\big]\,,\\
\Chi^{\dagger} &= -\frac{1}{2mc}\big[\bm{\sigma}\cdot \left(\hat{\bm{p}} -  q\bm{A}\right)\, \psi\big]^\dagger,
\end{align*}
which, when substituted in the last two terms of equation \eqref{ptrhoA1} involving $\bm{A}$ yield
\begin{align*}
    &-\frac{iq}{\hbar}\Big\{\psi^{\dagger}\big[\!\left(\bm{\sigma}\cdot\bm{A}\right)\Chi\big] - \big[\!\left(\bm{\sigma}\cdot \bm{A}\right)\Chi\big]^{\dagger}\psi\Big\} \\[5pt]
    &\quad=\frac{iq}{2mc\hbar}\Big\{\psi^{\dagger}\big[\!\left(\bm{\sigma}\cdot\bm{A}\right)\bm{\sigma}\cdot \left(\hat{\bm{p}} -  q\bm{A}\right)\psi\big] \\
    &\kern4cm -\big[\!\left(\bm{\sigma}\cdot\bm{A}\right)\bm{\sigma}\cdot \left(\hat{\bm{p}} -  q\bm{A}\right)\psi\big]^\dagger\psi\Big\} \\[5pt]
    &\quad= \frac{q}{2mc}\Big\{\psi^{\dagger}\big[\!\left(\bm{\sigma}\cdot\bm{A}\right)\!\left(\bm{\sigma}\cdot \grad \psi\right)\!\big]+ \big[\!\left(\bm{\sigma}\cdot\bm{A}\right)\!\left(\bm{\sigma}\cdot\grad\psi\right)^{\dagger}\big]\psi\Big\} \\
    &\qquad+\frac{iq^2}{2mc\hbar}\Big\{\psi^{\dagger}\big[\!\left(\bm{\sigma}\cdot\bm{A}\right)\!\left(\bm{\sigma}\cdot \bm{A} \psi\right)\!\big]\\
    &\kern4.7cm- \big[\!\left(\bm{\sigma}\cdot \bm{A}\right)\!\left(\bm{\sigma}\cdot \bm{A}\, \psi\right)^{\dagger}\!\big]\psi\Big\}\,.
\end{align*}
%
Here and in what follows we will make frequent use of the fact that $\bm{\sigma}\cdot \bm{A}$ is Hermitian, in which case the last two terms vanish, leaving
\begin{align*}
    &-\frac{iq}{\hbar}\Big\{\psi^{\dagger}\big[\!\left(\bm{\sigma}\cdot\bm{A}\right)\Chi\big] - \big[\!\left(\bm{\sigma}\cdot \bm{A}\right)\Chi\big]^{\dagger}\psi\Big\} \\[3pt]
    &\quad= \frac{q}{2mc}\Big\{\psi^{\dagger}\big[\!\left(\bm{\sigma}\cdot\bm{A}\right)\!\left(\bm{\sigma}\cdot \grad \psi\right)\!\big]+ \big[\!\left(\bm{\sigma}\cdot\bm{A}\right)\!\left(\bm{\sigma}\cdot\grad\psi\right)^{\dagger}\big]\psi\Big\}
\end{align*}
%
Substituting this result in Eq. \eqref{ptrhoA1} then gives
\begin{align}\label{ptrhoA2}
    \frac{1}{c}\frac{\partial \rho}{\partial t} &= \div (\psi^{\dagger}\bm{\sigma}\Chi + \Chi^{\dagger}\bm{\sigma}\psi) - \big[(\bm{\sigma} \cdot \grad \psi)^{\dagger}\Chi \nonumber\\
    &\quad + \Chi^{\dagger}(\bm{\sigma}\cdot \vnab \psi)\big]+\frac{q}{2mc}\Big\{\psi^{\dagger}\big[\!\left(\bm{\sigma}\cdot\bm{A}\right)\!\left(\bm{\sigma}\cdot\grad\psi\right)\!\big] \nonumber\\
    &\kern2.65cm+ \big[\!\left(\bm{\sigma}\cdot\bm{A}\right)\!\left(\bm{\sigma}\cdot \grad \psi\right)^{\dagger}\big]\psi\Big\}\,.
\end{align}
%
We next focus attention on the second term, in brackets, of Eq. \eqref{ptrhoA2}.  Substituting for $\Chi$ and $\Chi^\dagger$ in this term yields
\begin{align*}
    &(\bm{\sigma}\cdot\grad \psi)^{\dagger}\Chi + \Chi^{\dagger}(\bm{\sigma}\cdot \grad \psi) \\[5pt]
    &\quad=-\frac{1}{2mc}\Big\{(\bm{\sigma}\cdot\grad \psi)^{\dagger}\left[\bm{\sigma}\cdot \left(\hat{\bm{p}} -  q\bm{A}\right)\, \psi\right] \\
    &\kern3.65cm+ \left[\bm{\sigma}\cdot \left(\hat{\bm{p}} -  q\bm{A}\right)\, \psi\right]^{\dagger}(\bm{\sigma}\cdot \grad \psi)\Big\} \\[5pt]
    &\quad= \frac{i\hbar}{2mc}\Big\{(\bm{\sigma}\cdot \grad \psi)^{\dagger}\left(\bm{\sigma}\cdot \grad\, \psi\right)- \left(\bm{\sigma}\cdot \grad\, \psi\right)^{\dagger}(\bm{\sigma}\cdot \grad \psi)\Big\} \\
    &\qquad~~ +\frac{q}{2mc}\Big\{(\bm{\sigma}\cdot\grad \psi)^{\dagger}\left(\bm{\sigma}\cdot \bm{A}\, \psi\right) \\[-2pt]&\kern4.7cm+ \left(\bm{\sigma}\cdot \bm{A}\, \psi\right)^{\dagger}(\bm{\sigma}\cdot \grad \psi)\Big\}
\end{align*}
%
The first two terms on the right cancel, so we can substitute the last two in Eq. \eqref{ptrhoA2} to obtain an expression with two additional terms in braces:
\begin{align*}
    \frac{1}{c}\frac{\partial \rho}{\partial t} &= \div (\psi^{\dagger}\bm{\sigma}\Chi + \Chi^{\dagger}\bm{\sigma}\psi)+\frac{q}{2mc}\Big\{\psi^{\dagger}\!\big[\!\left(\bm{\sigma}\cdot\bm{A}\right)\!\left(\bm{\sigma}\cdot \grad\psi\right)\!\big]\\
    &\kern1cm+ \big[\!\left(\bm{\sigma}\cdot\bm{A}\right)\!\left(\bm{\sigma}\cdot \grad \psi\right)^{\dagger}\!\big]\psi - (\bm{\sigma}\cdot\grad \psi)^{\dagger}\!\left(\bm{\sigma}\cdot \bm{A}\psi\right)\\
    &\kern4.3cm + \left(\bm{\sigma}\cdot \bm{A} \psi\right)^{\dagger}\!(\bm{\sigma}\cdot \grad \psi)\Big\},
\end{align*}
%
and again, since $\bm{\sigma}\cdot \bm{A}$ is Hermitian, the terms involving $\bm{A}$ vanish, leaving simply
\begin{align*}
\frac{1}{c}\frac{\partial \rho}{\partial t} = \div (\psi^{\dagger}\bm{\sigma}\Chi + \Chi^{\dagger}\bm{\sigma}\psi),
\end{align*}
hence the current density can be identified as
\begin{align}
\bm{J} = -c\,(\psi^{\dagger}\bm{\sigma}\Chi + \Chi^{\dagger}\bm{\sigma}\psi)\,. \label{ptrhoA3}
\end{align}
Substituting once more for $\Chi$ and $\Chi^\dagger$, we obtain from Eq.~\eqref{ptrhoA3}:
\begin{align*}
\bm{J} &= \frac{1}{2m}\Big\{\psi^{\dagger}\bm{\sigma}\big[\bm{\sigma}\cdot (\hat{\bm{p}} -  q\bm{A}) \psi\big]+ \big[\bm{\sigma}\cdot (\hat{\bm{p}} -  q\bm{A}) \psi\big]^{\dagger}\!\bm{\sigma}\,\psi\Big\} \\
&=-\dfrac{i\hbar}{2m}\big[\psi^{\dagger}\bm{\sigma}(\bm{\sigma}\cdot \vnab \psi) - (\bm{\sigma}\cdot \vnab \psi)^\dagger\bm{\sigma}\psi\big]\\
&\quad\,- \frac{q}{2m}\big[\psi^{\dagger}\bm{\sigma}(\bm{\sigma}\cdot \bm{A} \psi) + (\bm{\sigma}\cdot \bm{A}\psi)^{\dagger}\bm{\sigma}\psi\big].
\end{align*}
%
From the derivation leading to Eqs \eqref{J1} and \eqref{Jtotal}, the first two terms in braces can be replaced to yield
\begin{align*}
    \bm{J} &= -\dfrac{i\hbar}{2m}\big[\psi^{\dagger}\grad\psi - (\vnab \psi)^{\dagger}\psi\big]+ \dfrac{\hbar}{2m}\curl (\psi^{\dagger}\bm{\sigma}\psi)\\&\quad\,- \frac{q}{2m}\big[\psi^{\dagger}\bm{\sigma}(\bm{\sigma}\cdot \bm{A} \psi) +  \psi^{\dagger}(\bm{\sigma}\cdot \bm{A})\bm{\sigma}\psi\big].
\end{align*}
%
The $i$th component of the terms involving $\bm{A}$ can be written, making use of the Pauli matrix identity $\sigma_i\sigma_j + \sigma_j\sigma_i = 2\delta_{ij}$, very simply as
\begin{align*}
    \psi^{\dagger}\sigma_i(\sigma_j A_j \psi) + \psi^{\dagger}(\sigma_j A_j)\sigma_i\psi&=\psi^{\dagger}(\sigma_i\sigma_j + \sigma_j\sigma_i) A_j \psi \\&= 2\psi^{\dagger}\delta_{ij}A_j\psi = 2 A_i\psi^{\dagger}\psi,
\end{align*}
%
which, when substituted in the previous equation, yields the correct non-relativistic current density for a spin-1/2 particle interacting with an external electromagnetic field:
\begin{align}
    \bm{J} &= -\dfrac{i\hbar}{2m}\big[\psi^{\dagger}\grad\psi - (\vnab \psi)^{\dagger}\psi\big] + \dfrac{\hbar}{2m}\curl(\psi^{\dagger}\bm{\sigma}\psi) \nonumber\\&\quad\, - \frac{q}{m}\bm{A}\psi^{\dagger}\psi, \label{JtotalA}
\end{align}
%
verifying Eqs. \eqref{DiracNRJ} and \eqref{JNRspin}.

\section{Does the Spin Current Have Measurable Consequences?}

Although the spin current has no effect in the {\emph{probability conservation equation}}, it has been convincingly demonstrated to be an inherent feature of the {\emph{probability current density}} of a spin-1/2 particle described by the \SP equation. If the particle carries a charge $q$, then it should contribute to the {\emph{charge current density}}, hence be measurable in experiments sensitive to the effects of such currents. After all, paraphrasing Nowakowski in Ref.\ \onlinecite{Nowakowski}, electric current is a physical observable.

In Ref.\ \onlinecite{Nowakowski} an interesting experiment is proposed and analyzed to demonstrate the relevance of the spin current. It is determined that in such an experiment there would be a contribution to the electric current density ``which is of purely quantum mechanical origin and which can be traced back to the `spin term' $\dots$ .'' The author further speculates that ``this term might play a role in problems concerning conductivity in solids.''

Du Toit, in his Honors Thesis \cite{Dutoit}, discusses a simpler experiment, an electron in a homogeneous magnetic field (the Landau problem). His analysis is rather inconclusive, the main effect being described as a ``swirl'' of charge current creating a current loop/dipole moment that could possibly interact with the field to produce a torque on the dipole, presumably observable as some sort of precessional motion.

A field of physics where the current density plays a {\emph {critical}} role is Bohmian mechanics (or de Broglie-Bohm theory; see Refs. \onlinecite{Durr}--\onlinecite{BohmHiley} for excellent general accounts). Well-defined concepts of point particles and trajectories are fundamental features of this theory, where particle motion takes place along trajectories that are integral curves of a velocity field proportional to the current density (Ref.\ \onlinecite{BohmHiley}, Ch. 10). Thus, a particle's equations of motion are directly determined by the current density itself, hence the spin term can directly influence its motion. In recent work \cite{Sid1, Sid2} directed toward the prediction of arrival times in time-of-flight experiments, the possibility of observing rather remarkable effects of the spin term is discussed. In these papers, time-of-flight experiments are proposed and carefully analyzed for electrons prepared in various initial states, moving in a cylindrical waveguide. Analytical and numerical methods are used to obtain predictions of their arrival time distributions at a detector. For a particular initial state, with spin vector oriented perpendicular to the waveguide axis, the predicted distributions show a striking result, a cutoff, or maximum arrival time, after which no further electrons would be detected. Such an experiment, or one similar to it, seems to be well within the scope of present day technology. The results would be a welcome test of the ability of Bohmian mechanics to predict arrival times of spin-1/2 particles, given that a standard quantum mechanical prediction of the same is still ambiguous \cite{Muga2000}.

Finally, as further examples of the possible impact of the spin term, we mention recent publications \cite{BB1, BB2} applying the Dirac theory to experiments with relativistic electron beams. In these, a velocity field of probability flow is defined by precisely the same equation used in Bohmian mechanics to describe particle trajectories, i.e., proportional to the current density, and used to describe results of such experiments (or suggested ones). As in the non-relativistic case, the spin term makes a definite contribution that cannot be ignored.

\section{Concluding Remarks}

After reviewing previous work, our goals here were two-fold.\ First, to emphasize that although the Pauli equation can be {\emph {constructed}} from the \Sch equation by judicious use of spin operator identities, thus demonstrating the non-relativistic nature of the electron spin, \LL pointed the way to a direct {\emph {derivation}} of the Pauli equation via a linearization of the \Sch equation, from which the non-relativistic nature of the electron spin naturally follows.\ Second, to direct attention to the work of Shikakhwa, et al.\ \cite{Shikakhwa}, who showed conclusively and unambiguously that the spin term of the probability current density is a non-relativistic phenomenon derivable from the \SP equation, contrary to what is sometimes claimed in the literature \cite{Nowakowski}. Furthermore, we emphasized the importance of the \LL theory in this regard, showing that the spin term is directly derivable from his equation, which is properly viewed as a precursor of the Pauli equation.

We conclude by cautioning readers about the casual use of the ubiquitous minimal coupling prescription in this and other work, invoking electromagnetic potentials that are inherently relativistic (since they originate in the Lorentz covariant Maxwell equations), and using them in a non-relativistic quantum theory involving the \Schn, Pauli, or \LL equations.\ Constraints on the allowable electromagnetic fields as a result of requiring Galilean covariance of both the quantum theory and electromagnetism were first discussed by \LL in Ref. \onlinecite{Levyleblond1}, p. 305. His results have been sharpened and developed considerably in later work \cite{LeBellac, BrownHolland, HollandBrown, deMontigny1, deMontigny2}, but a discussion of these important articles is beyond the scope of the present paper. A review of this work is highly recommended to anyone applying non-relativistic quantum theories to the analysis of experimental results involving electromagnetic interactions.

\begin{acknowledgments}

My thanks to Siddhant Das for his encouragement in pursuing a better understanding of the Pauli equation and its spin current density, and for critically reviewing the manuscript. I am also grateful to Detlef D{\"{u}}rr for his interest in this work, and for his collegial grace in making me feel welcome as a long-distance collaborator on research efforts at the Mathematisches Institut, Ludwig-Maximilians-Universitat M{\"{u}}nchen.

\end{acknowledgments}

\end{document}